\documentclass[english,prl, twocolumn]{revtex4-1}
\usepackage[T1]{fontenc}
\usepackage[latin9]{inputenc}
\usepackage{amstext}
\usepackage{graphicx}
\usepackage{babel}
\begin{document}

\title{A simple, general result for the variance of substitution number
in molecular evolution. }

\author{Bahram Houchmandzadeh and Marcel Vallade}

\address{CNRS, LIPHY, F-38000 Grenoble, France\\
Univ. Grenoble Alpes, LIPHY, F-38000 Grenoble, France}
\begin{abstract}
The number of substitutions (of nucleotides, amino acids, ...) that
take place during the evolution of a sequence is a stochastic variable
of fundamental importance in the field of molecular evolution. Although
the \emph{mean} number of substitutions during molecular evolution
of a sequence can be estimated for a given substitution model, no
simple solution exists for the \emph{variance} of this random variable. 

We show in this article that the computation of the variance is as
simple as that of the mean number of substitutions for both short
and long times. Apart from its fundamental importance, this result
can be used to investigate the dispersion index $R$, \emph{i.e.}
the ratio of the variance to the mean substitution number, which is
of prime importance in the neutral theory of molecular evolution.
By investigating large classes of substitution models, we demonstrate
that although $R\ge1$, to obtain $R$ significantly larger than unity
necessitates in general additional hypotheses on the structure of
the substitution model.

\end{abstract}
\maketitle

\section{Introduction.}

Evolution at the molecular level is the process by which random mutations
change the content of some sites of a given sequence (of nucleotides,
amino acids, ...) during time. The number of substitutions $n$ that
occur during this time is of prime importance in the field of molecular
evolution and its characterization is the first step in deciphering
the history of evolution and its many branching. The main observable
in molecular evolution, on comparing two sequences, is $\hat{p}$,
the fraction of sites at which the two sequences are different. In
order to estimate the statistical moments of $n$, the usual approach
is to postulate a substitution model $\mathbf{Q}$ through which $\hat{p}$
can be related to the statistical moments of $n$. The simplest and
most widely used models assume that $\mathbf{Q}$ is site independent,
although this constraint can be relaxed\cite{Graur1999,Yang2006}.

Once a substitution model $\mathbf{Q}$ has been specified, it is
straightforward to deduce the \emph{mean} number of substitutions
$\left\langle n\right\rangle $ and the process is detailed in many
textbooks. However, the mean is only the first step in the characterization
of a random variable and by itself is a rather poor indicator. The
next step in the investigation of a random variable is to obtain its
variance $V$. Surprisingly, no simple expression for $V$ can be
found in the literature for arbitrary substitution model $\mathbf{Q}$.
The first purpose of this article is to overcome this shortcoming.
We show that computing $V$ is as simple as computing $\left\langle n\right\rangle $,
both for short and long times.

We then apply this fundamental result to the investigation of the
dispersion index $R$, the ratio of the variance to the mean number
of substitutions. The neutral theory of molecular evolution introduced
by Kimura\cite{Kimura1984} supposes that the majority of mutations
are neutral (\emph{i.e. }have no effect on the phenotypic fitness)
and therefore substitutions in protein or DNA sequences accumulate
at a ``constant rate'' during evolution, a hypothesis that plays
an important role in the foundation of the ``molecular clock''\cite{Bromham2003,Ho2014}.
The original neutral theory postulated that the substitution process
is Poissonian, \emph{i.e.} assuming $R=1$. Since the earliest work
on the index of dispersion, it became evident however that $R$ is
usually much larger than unity (see \cite{Cutler2000a} for a review
of data). Many alternatives have been suggested to reconcile the ``overdispersion''
observation with the neutral theory (\cite{Cutler2000a}). Among these
various models, a promising alternative, that of fluctuating neutral
space, was suggested by Takahata \cite{Takahata1991a} which has been
extensively studied in various frameworks (\cite{Zheng2001,Bastolla2002,Wilke2004,Bloom2007,Raval2007}). 

The fluctuating neutral space model states that the substitution rate
$m_{j}^{i}$ from state $i$ to state $j$ is a function of both $i$
and $j$. States $i$ and $j$ can be nucleotides or amino acids,
in which case we recover the usual substitution models of molecular
evolution discussed above. The states can also be nodes of a neutral
graph used to study global protein evolution (\cite{Huynen1996,Bornberg-Bauer1999,VanNimwegen1999}).
For neutral networks used in the study of protein evolution, Bloom,
Raval and Wilke \cite{Bloom2007} devised an elegant procedure to
estimate the substitution rates. We will show in this article that
in general $R\ge1$ and the equality is reached only for the most
trivial cases. However, producing large $R$ requires additional hypotheses
on the structure of substitution rates. 

In summary, the problem we investigate in this article is to find
a simple and general solution for the variance and dispersion index
of any substitution matrix of dimension $K$. A substitution matrix
$\mathbf{Q}$ collects the transition rates $m_{j}^{i}$ ($i\ne j$);
its diagonal elements $q_{i}^{i}=-m^{i}$ are set such that its columns
sum to zero (see below for notations) and designate the rate of leaving
state $i$. Because of this condition, $\mathbf{Q}$ is singular. 

Zheng \cite{Zheng2001} was the first to use Markov chains to investigate
the variance of substitution number as a solution of a set of differential
equations. His investigation was further developed by Bloom, Raval
and Wilke \cite{Bloom2007} who gave the general solution in terms
of the spectral decomposition of the substitution matrix; this solution
was extended by Raval \cite{Raval2007} for a specific class of matrices
used for random walk on neutral graphs. Minin and Suchard \cite{Minin2008}
used the same spectral method to derive an analytical form for the
generating function of a binary process.

The first step to characterize the substitution number, which as is
well known, is to find the equilibrium probabilities $\pi_{i}$ of
being in a state $i$, which is obtained by solving the linear system
$\sum_{i}q_{j}^{i}\pi_{i}=0$ with the additional condition of $\sum_{i}\pi_{i}=1$.
Once $\pi_{i}$ are obtained, the \emph{mean} substitution number
as a function of time is simply $\left\langle n\right\rangle =\bar{m}t$
where $\bar{m}=\sum_{i}m^{i}\pi_{i}$ is the weighted average of the
``leaving'' rates. 

We show here that finding the variance necessitates a similar computation.
Denoting the weighted deviation of the diagonal elements of $\mathbf{Q}$
from the mean $h_{i}=(\bar{m}-m^{i})\pi_{i}$, we have to find the
solution of the linear system $\sum_{i}q_{j}^{i}r_{i}=h_{j}$ with
the additional condition $\sum r_{i}=0$. For long times, the dispersion
index is then simply 
\begin{equation}
R=1+\frac{2}{\bar{m}}\sum_{i=1}^{K}m^{i}r_{i}\label{eq:mainequation}
\end{equation}

For short times, \emph{i.e. }when the mean number of substitutions
is small, the result is even simpler : 
\begin{equation}
R=1+\frac{v_{m}}{\bar{m}^{2}}\left\langle n\right\rangle \label{eq:mainequationb}
\end{equation}
where 
\[
v_{m}=\sum_{i=1}^{K}(\bar{m}-m^{i})^{2}\pi_{i}
\]
in other words, $v_{m}$ is the variance of the diagonal elements
of the substitution matrix, weighted by the equilibrium probabilities. 

This article is organized as follow. In the next section, we use a
Markov chain approach to derive relations (\ref{eq:mainequation},\ref{eq:mainequationb})
and show its validity by comparing it to results obtained by direct
numerical simulations. The simplicity of these results then allows
us to study the dispersion index for specific models of nucleotide
substitutions widely used in the literature (section \ref{sec:specificmodel})
and for general models (section \ref{sec:generalmodel}). We investigate
in particular the conditions necessary to produce large $R$. The
last section is devoted to a general discussion of these results and
to conclusions. Technical details, such as the proof of $R\ge1$ are
given in the appendices.

\section{Markov chain model of dispersion index.}

\subsection{Background and definitions.}

The problem we investigate in this article is mainly that of counting
transitions of a random variable
\begin{figure}
\begin{centering}
\includegraphics[width=0.6\columnwidth]{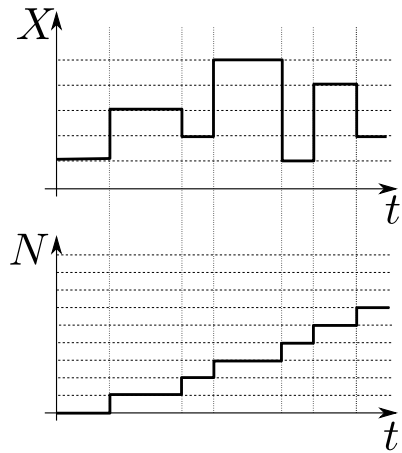}
\par\end{centering}

\caption{The random variables $X$ can switch between $K$ states; the counter
$N$ of the number of transitions is incremented at each transition
of the variable $X$. The figure above shows one realization of these
random variables as a function of time.\label{fig:countingX}}

\end{figure}
 (figure \ref{fig:countingX}). Consider a random variable $X$ that
can occupy $K$ distinct states and let $m_{j}^{i}$ ($i\ne j$, $1\le i,j\le K$
) be the transition rate from state $i$ to state $j$. The probability
density $p_{i}(t)$ of being in state $i$ at time $t$ is governed
by the Master equation 
\begin{eqnarray*}
\frac{dp_{i}}{dt} & = & -\sum_{j}m_{j}^{i}p_{i}+\sum_{j}m_{i}^{j}p_{j}\\
 & = & -m^{i}p_{i}+\sum_{j}m_{i}^{j}p_{j}
\end{eqnarray*}
where $m^{i}=\sum_{j}m_{j}^{i}$ is the ``leaving'' rate from state
$i$. We can collect the $p_{i}$ into a (column) vector $\left|p\right\rangle =(p_{1},...p_{K})^{T}$
and write the above equations in matrix notation
\begin{equation}
\frac{d}{dt}\left|p\right>=(-\mathbf{D}+\mathbf{M})\left|p\right>=\mathbf{Q}\left|p\right>\label{eq:mainmasterP}
\end{equation}
where $\mathbf{D}$ is the diagonal matrix of $m^{i}$ and $\mathbf{M}=(m_{j}^{i})$
collects the detailed transition rates from state $i$ to state $j$
($i\ne j$) and has zero on its diagonal. In our notations, the upper
(lower) index designates the column (row) of a matrix. The matrix
$\mathbf{Q}=-\mathbf{D}+\mathbf{M}$ is called the substitution matrix
and its columns sum to zero.

Before proceeding, we explain the notations used in this article.
As the matrix $\mathbf{Q}$ is not in general symmetric, a clear distinction
must be made between right (column) and left (row) vectors. The Dirac
notations are standard and useful for handling this distinction :
a column vector $(x_{1},...x_{K})^{T}$ is denoted $\left|x\right>$
while a row vector $(y^{1},...,y^{K})$ is denoted $\left<y\right|$
and $\left\langle y|x\right\rangle =\sum_{i}y^{i}x_{i}$ is their
scalar product. In some of literature (see \cite{Yang2006}), the
substitution matrix is the transpose of the matrix used here and the
master equation is then written as $d\left<p\right|/dt=\left<p\right|\mathbf{Q}$
and therefore its rows sum to zero. 

By construction, the matrix $\mathbf{Q}$ is singular and has one
zero eigenvalue while all others are negative. Therefore, as time
flows, $\left|p(t)\right>\rightarrow\left|\pi\right>$ where $\left|\pi\right>=(\pi_{1},...\pi_{K})$
is the equilibrium occupation probability and the zero-eigenvector
of the substitution matrix. 
\begin{eqnarray*}
\mathbf{Q}\left|\pi\right> & = & 0\\
\left\langle 1|\pi\right\rangle  & = & 1
\end{eqnarray*}
where $\left<1\right|=(1,...1)$, and the second condition expresses
that the sum of the probabilities must be 1. Note that by definition,
$\left<1\right|\mathbf{Q}=0$, and thus $\left<1\right|$ is a zero
left eigenvector of the substitution matrix.

\subsection{Problem formulation.}

To count the number of substitutions (figure \ref{fig:countingX}),
we consider the probability densities $p_{i}^{n}(t)$ of being in
state $i$ after $n$ substitutions at time $t$. These probabilities
are governed by the master equation 
\begin{eqnarray*}
\frac{dp_{i}^{n}}{dt} & = & -m^{i}p_{i}^{n}+\sum_{j}m_{i}^{j}p_{j}^{n-1}\,\,\,\,\,\,n>0\\
\frac{dp_{i}^{0}}{dt} & = & -m^{i}p_{i}^{0}
\end{eqnarray*}
We can combine the above equations by setting $p_{i}^{n}(t)=0$ if
$n<0$. Collecting the elements of $(p_{1}^{n},p_{2}^{n},...,p_{K}^{n})^{T}$
into the vector $\left|p^{n}\right>$, the above equation can then
be written as 
\begin{equation}
\frac{d}{dt}\left|p^{n}\right>=-\mathbf{D}\left|p^{n}\right>+\mathbf{M}\left|p^{n-1}\right>\label{eq:MasterN}
\end{equation}

The quantities of interest for the computation of the dispersion index
are the mean and the variance of the number of substitutions. The
mean number of substitutions at time $t$ is 
\[
\left\langle n(t)\right\rangle =\sum_{i,n}np_{i}^{n}(t)
\]
Let us define 
\[
n_{i}(t)=\sum_{n}np_{i}^{n}(t)
\]
and collect the partial means $n_{i}$ into the vector $\left|n(t)\right>=(n_{1},...,n_{K})^{T}$.
The mean is then defined simply as 
\[
\left\langle n(t)\right\rangle =\sum_{i}n_{i}(t)=\left\langle 1|n(t)\right\rangle 
\]
By the same token, the second moment 
\[
\left\langle n^{2}(t)\right\rangle =\sum_{i,n}n^{2}p_{i}^{n}(t)
\]
can be written in terms of partial second moments $n_{i}^{2}=\sum_{n}n^{2}p_{i}^{n}(t)$
as 
\[
\left\langle n^{2}(t)\right\rangle =\left\langle 1|n^{2}(t)\right\rangle 
\]
where $\left|n^{2}(t)\right>=(n_{1}^{2},...,n_{K}^{2})^{T}$. It is
straightforward to show (see appendix \ref{sec:Mean-and-variance}),
for the initial condition $\left|p^{n}(0)\right\rangle =\left|\pi\right\rangle $,
that $\left|n(t)\right>$ and $\left|n^{2}(t)\right>$ obey a linear
differential equation 
\begin{eqnarray}
\frac{d}{dt}\left|n\right> & = & \mathbf{Q}\left|n\right>+\mathbf{D}\left|\pi\right>\label{eq:moment1}\\
\frac{d}{dt}\left|n^{2}\right> & = & \mathbf{Q}\left|n^{2}\right>+2\mathbf{M}\left|n\right>+\mathbf{D}\left|\pi\right>\label{eq:moment2}
\end{eqnarray}
Let us define the left vector 
\[
\left<m\right|=(m^{1},...,m^{K})
\]
which collects the leaving rates. By definition, $\left<1\right|\mathbf{M}=\left<1\right|\mathbf{D}=\left<m\right|$.
Multiplying eqs (\ref{eq:moment1},\ref{eq:moment2}) by the left
vector $\left<1\right|$, and noting that $\left<1\right|\mathbf{Q}=\left<0\right|$,
we get a simple relation for the moments : 
\begin{eqnarray}
\frac{d}{dt}\left\langle n\right\rangle  & = & \left\langle m|\pi\right\rangle \label{eq:nmoment1}\\
\frac{d}{dt}\left\langle n^{2}\right\rangle  & = & 2\left\langle m|n\right\rangle +\left\langle m|\pi\right\rangle \label{eq:nmoment2}
\end{eqnarray}
We observe that the mean number of substitutions involves only a trivial
integration. Defining the weighted average of the leaving rates as
\[
\bar{m}=\left\langle m|\pi\right\rangle =\sum_{i}m^{i}\pi_{i}
\]
the mean number of substitution is simply 
\begin{equation}
\left\langle n(t)\right\rangle =\bar{m}t\label{eq:mean1}
\end{equation}

To compute the second moment of the substitution number on the other
hand, we must solve for $\left|n\right>$ using equation (\ref{eq:moment1})
and then perform one integration. The next subsection is devoted to
the efficient solution of this procedure.

\subsection{Solution of the equation for the moments . \label{sub:Solution}}

One standard way of solving equation (\ref{eq:moment1}) would be
to express the matrix $\mathbf{Q}$ in its eigenbasis; equation(\ref{eq:moment1})
is then diagonalized and can be formally solved. This is the method
used by Bloom, Raval and Wilke \cite{Bloom2007} and further refined
by Raval\cite{Raval2007} for a specific class of substitution matrices
where $m_{j}^{i}=0\,\mbox{or}\,1$. The first problem with this approach
is that there is no guarantee that $\mathbf{Q}$ is diagonalizable.
Even if $\mathbf{Q}$ can be diagonalized, this is not the most efficient
procedure to find $V$, as it necessitates the computation of all
eigenvalues and left and right eigenvectors of $\mathbf{Q}$ and then
the cumbersome summation of their binomial products. 

The procedure we follow involves some straightforward, albeit cumbersome
linear algebraic operations, but the end result is quite simple. We
note that the matrix $\mathbf{Q}$ is singular and has exactly one
zero eigenvalue, associated with the left $\left<1\right|$ and right
$\left|\pi\right>$ eigenvectors. The method we use is to isolate
the zero eigenvalue by making a round-trip to a new basis. Thus, if
we can find a new basis in which the substitution matrix $\mathbf{Q}'=\mathbf{X}^{-1}\mathbf{Q}\mathbf{X}$
takes a lower block triangular form

\begin{equation} \label{Q'}
\mathbf{Q}'=\left( 
\begin{array}{c|c} 
0  & 0 \cdots 0 \\ \hline   
\  & \raisebox{-3ex}{{\large\mbox{{$\mathbf{\tilde{Q}}$}}}} \\[-3.3ex]  
\tilde{\alpha} & \\[-0.5ex]   \  & 
\end{array} 
\right)  			
\end{equation}we will have  achieved our goal of isolating the zero eigenvalue.
The non singular matrix $\mathbf{\tilde{\mathbf{Q}}}$ is of rank
$K-1$ and has the non-zero and negative eigenvalues of $\mathbf{Q}$.
As $\left<1\right|$ is the known left eigenvalue of $\mathbf{Q}$,
we can split the vector space into ${\cal B}=\{\left|u\right\rangle $|
$\left\langle 1|u\right\rangle =0\}$ and the space padded by $\left|\pi\right\rangle $.
It is then straightforward to find the above transfer matrices $\mathbf{X}$
and $\mathbf{X}^{-1}$ for such a transformation:

\[
\mathbf{X}=\left(\begin{array}{ccccc}
1 & -1 & -1 & \cdots & -1\\
0 & 1 & 0 & \cdots & 0\\
0 & 0 & 1 & \cdots & 0\\
\vdots &  &  & \ddots & \vdots\\
0 &  & \cdots & 0 & 1
\end{array}\right);\mathbf{X}^{-1}=\left(\begin{array}{ccccc}
1 & 1 & 1 & \cdots & 1\\
0 & 1 & 0 & \cdots & 0\\
0 & 0 & 1 & \cdots & 0\\
\vdots &  &  & \ddots & \vdots\\
0 &  & \cdots & 0 & 1
\end{array}\right)
\]
Under such a transformation, a right vector $\left|x\right\rangle =(x_{1},x_{2},...,x_{K})^{T}$
transforms into 
\[
\left|x'\right\rangle =\mathbf{X}^{-1}\left|x\right\rangle =\left(\begin{array}{c}
\sum_{i}x_{i}\\
x_{2}\\
\vdots\\
x_{K}
\end{array}\right)=\left(\begin{array}{c}
\sum_{i}x_{i}\vspace{0.5ex}\\
\hline \\
\left|\tilde{x}\right\rangle \\
\\
\end{array}\right)
\]
where the $K-1$ dimensional vector $\left|\tilde{x}\right\rangle =(x_{2},...,x_{K})^{T}$.
In general, we will designate by a $\tilde{}$ ~all vectors that
belong the $K-1$ dimensional space ${\cal B}$ in which the linear
application $\mathbf{\tilde{Q}}$ operates. 

A left vector $\left<y\right|=(y^{1},y^{2},...,y^{K})$ transforms
into 
\[
\left<y'\right|=\left<y\right|\mathbf{X}=(y^{1},y^{2}-y^{1},...,y^{K}-y^{1})=(y^{1}\vline\ \left<\tilde{y}\right|\ )
\]
where the $K-1$ dimensional left vector $\left<\tilde{y}\right|=(y^{2}-y^{1},...,y^{K}-y^{1})$.

Finally, $\tilde{Q}_{j}^{i}=Q_{j}^{i}-Q_{j}^{1}$ where the elements
of $\tilde{\mathbf{Q}}$ have been indexed from 2 to $K$. 

Expressing now the equation (\ref{eq:moment1}) for the evolution
of first moments in the new basis, we find that 
\begin{eqnarray}
\frac{d}{dt}\left\langle n\right\rangle  & = & \bar{m}\label{eq:tilde0}\\
\frac{d}{dt}\left|\tilde{n}\right\rangle  & = & \mathbf{\tilde{Q}}\left|\tilde{n}\right\rangle +\left\langle n\right\rangle \left|\tilde{\alpha}\right\rangle +\left|\tilde{\mu}\right\rangle \label{eq:tilde1}
\end{eqnarray}
where $\left|\tilde{n}\right\rangle =(n_{2},...,n_{K})^{T}$, $\left|\tilde{\mu}\right\rangle =(m^{2}\pi_{2},...,m^{K}\pi_{K})$
and $\left|\tilde{\alpha}\right\rangle $ is given in relation \ref{Q'}.
Equation (\ref{eq:tilde0}) is the same as equation (\ref{eq:nmoment1})
and implies that $\left\langle n\right\rangle =\bar{m}t$. As $\tilde{\mathbf{Q}}$
is non-singular (and negative definite), equations (\ref{eq:tilde0},\ref{eq:tilde1})
can now readily be solved. Noting that $\mathbf{Q}\left|\pi\right\rangle =0$
implies that $\left|\tilde{\alpha}\right\rangle +\tilde{\mathbf{Q}}\left|\tilde{\pi}\right\rangle =0$,
the differential equation (\ref{eq:tilde1}) integrates 
\begin{equation}
\left|\tilde{n}\right\rangle =\left(\mathbf{I}-e^{\tilde{\mathbf{Q}}t}\right)\tilde{\mathbf{Q}}^{-1}\left|\tilde{h}\right\rangle +\left\langle n\right\rangle \left|\tilde{\pi}\right\rangle \label{eq:ntilde}
\end{equation}
where $\left|\tilde{h}\right\rangle =\bar{m}\left|\tilde{\pi}\right\rangle -\left|\tilde{\mu}\right\rangle $. 

To compute the second moment (equation \ref{eq:nmoment2}) and the
variance, we need must integrate the above expression one more time.
We finally obtain
\begin{eqnarray}
\mbox{Var}(n) & = & \left\langle n^{2}\right\rangle -\left\langle n\right\rangle ^{2}\label{eq:variance}\\
 & = & \left\langle n\right\rangle +2\left\langle \tilde{m}|\left(It+\tilde{\mathbf{Q}}^{-1}(\mathbf{I}-e^{\tilde{\mathbf{Q}}t})\right)\tilde{\mathbf{Q}}^{-1}|\tilde{h}\right\rangle \nonumber 
\end{eqnarray}
The second term in the r.h.s. of the above equation is the excess
variance $\delta V$ with respect to a Poisson process.

\subsubsection{Long time behavior.}

As all eigenvalues of $\tilde{\mathbf{Q}}$ are negative, for large
times $\exp(\tilde{\mathbf{Q}}t)\rightarrow0$ and the leading term
of the excess variance is therefore 
\begin{equation}
\delta V=2\left\langle \tilde{m}|\tilde{r}\right\rangle t\label{eq:deltaVtilde}
\end{equation}
 where $\left|\tilde{r}\right\rangle $ is the solution of the linear
equation 
\begin{equation}
\tilde{\mathbf{Q}}\left|\tilde{r}\right\rangle =\left|\tilde{h}\right\rangle \label{eq:tilder}
\end{equation}
Returning to the original basis, relation (\ref{eq:deltaVtilde})
becomes 
\begin{equation}
\delta V=2\left\langle m|r\right\rangle t\label{eq:deltaVlong}
\end{equation}
where $\left<m\right|=(m^{1},m^{2},...m^{K})$ is the left vector
of the leaving rates and $\left|r\right\rangle $ is the solution
of the linear equation 
\begin{eqnarray}
\mathbf{Q}\left|r\right\rangle  & = & \left|h\right\rangle \label{eq:r1}\\
\left\langle 1|r\right\rangle  & = & 0\label{eq:r2}
\end{eqnarray}
$\left|h\right\rangle =(h_{1},...,h_{K})^{T}$ is the vector of weighted
deviation from $\bar{m}$ of the leaving rates $m^{i}$: 
\[
h_{i}=(\bar{m}-m^{i})\pi_{i}
\]
Finally, for large times, the dispersion index is 
\begin{equation}
R=1+2\frac{\left\langle m|r\right\rangle }{\bar{m}}\label{eq:dispersion}
\end{equation}
which is the relation (\ref{eq:mainequation}) given in the introduction. 

\begin{figure}
\begin{centering}
\includegraphics[width=0.8\columnwidth]{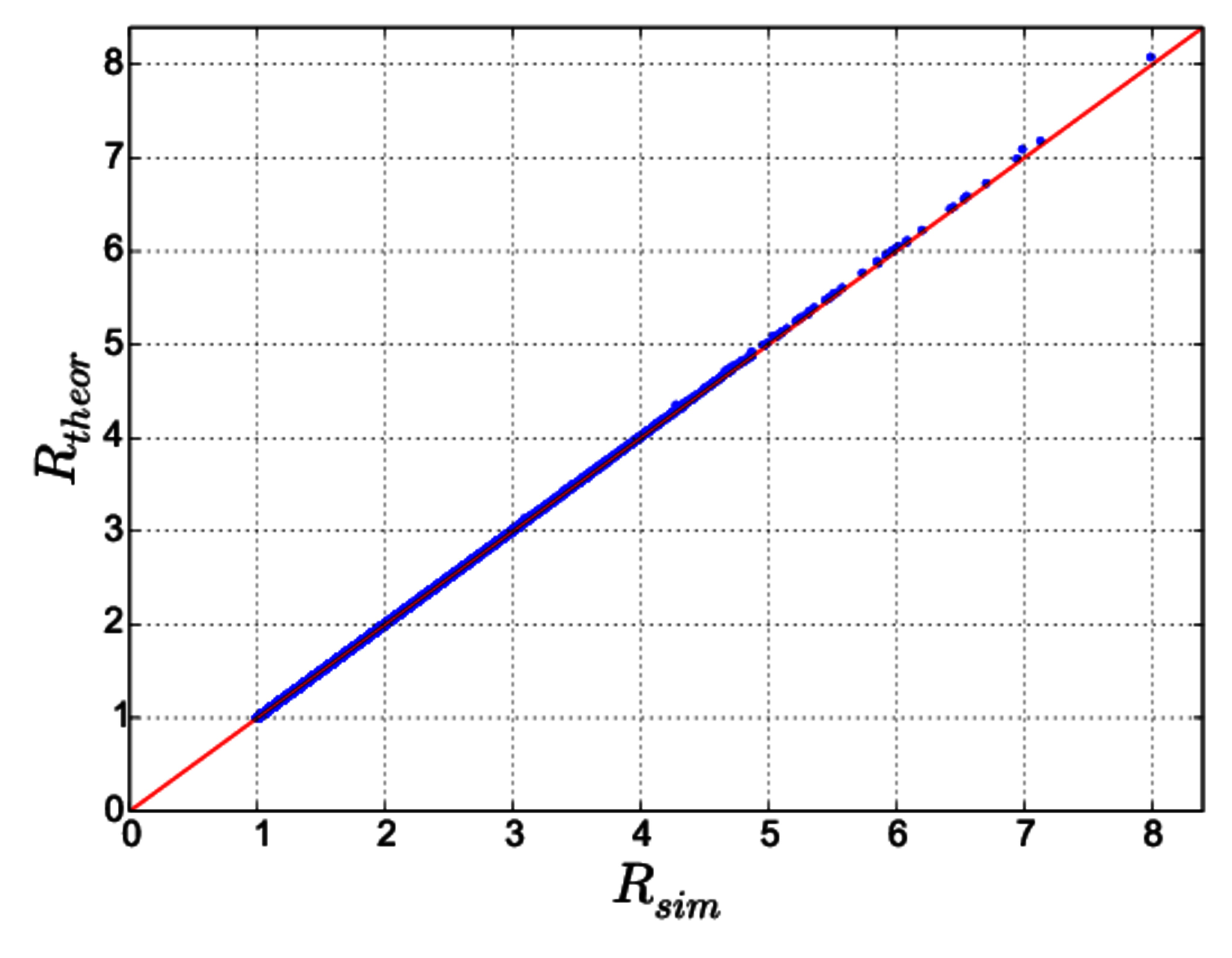}
\par\end{centering}

\caption{Comparison between the theoretical result (\ref{eq:dispersion}) and
numerical simulations. $3\times10^{5}$ $4\times4$ random (uniform$(0,1)$
) matrices were generated. For each matrix, a Gillespie algorithm
was used to generate $10^{6}$ random paths as a function of time
($t_{final}=1000$), from which the dispersion index was computed.
In the above figure, each dot corresponds to one random matrix. The
mean relative error $(R_{\mbox{theor}}-R_{\mbox{sim}})/R_{\mbox{theor}}$
is $1.3\times10^{-3}$. \label{fig:ThSim}}

\end{figure}

Figure \ref{fig:ThSim} shows the agreement between the above theoretical
results and stochastic numerical simulations.

Two important consequences should be noted. First, it is not difficult
to show that $R\ge1$, which is called the overdispersion of the molecular
clock. A demonstration of this theorem for symmetric substitution
matrices whose elements are 0 or 1 (adjacency matrices) was given
by Raval \cite{Raval2007}. We give the general demonstration for
general time reversible (GTR) substitution matrices in appendix \ref{sec:RL1}
. 

The second consequence of relation (\ref{eq:dispersion}) is that
if all diagonal elements of the substitution matrix are equal (\emph{i.e.
}$m^{i}=m^{j}$ $\forall i,j$ ), then the dispersion index is exactly
1 and we recover the property of a normal Poisson process, regardless
of the fine structure of $\mathbf{Q}$ and the equilibrium probabilities
$\pi_{i}$. This is a sufficient condition. We show that the necessary
condition for $R=1$ is $\left|h\right\rangle =\left|0\right\rangle $,
which, except for the trivial case where some $\pi_{i}=0$, again
implies the equality of diagonal elements of $\mathbf{Q}$ (see \ref{sec:RL1}).

\subsubsection{Short time behavior.\label{sub:Short-time-behavior}}

For short times, \emph{i.e.} when the mean number of substitution
is small, we can expand $\delta V$ given by expression (\ref{eq:variance})
to the second order in time: 
\begin{eqnarray}
\delta V & = & -t^{2}\left\langle \tilde{m}|\tilde{h}\right\rangle +O(t^{3})\label{eq:shorttime1}\\
 & = & -t^{2}\sum_{i=2}^{K}(m^{i}-m^{1})(\bar{m}-m^{i})\pi_{i}+O(t^{3})\label{eq:shorttime2}
\end{eqnarray}
Note that the above summation is over $i=2\cdots K$ . However, by
definition, 
\[
\sum_{i=1}^{K}(\bar{m}-m^{i})\pi_{i}=0
\]
and hence the sum in relation (\ref{eq:shorttime2}) can be rearranged
as 
\begin{equation}
\delta V=t^{2}\sum_{i=1}^{K}(\bar{m}-m^{i})^{2}\pi_{i}\label{eq:shortterm3}
\end{equation}
The sum,which we will denote by $v_{m}$, represents the variance
of the diagonal elements of $\mathbf{Q}$, weighted by the equilibrium
probabilities. It is more meaningful to express the variance in terms
of the mean substitution number. Using relation (\ref{eq:mean1}),
we therefore have 
\begin{equation}
\delta V=\frac{v_{m}}{\bar{m}^{2}}\left\langle n\right\rangle ^{2}\label{eq:shorttimevar}
\end{equation}
The dispersion index for short times is therefore 
\[
R=1+\frac{v_{m}}{\bar{m}^{2}}\left\langle n\right\rangle 
\]

The dispersion index for all times can also be computed, and an example
is given in appendix \ref{sec:alltimes}. In the next section, we
investigate some applications of the relation (\ref{eq:dispersion}).

\section{Application to specific nucleotides substitution models.\label{sec:specificmodel}}

Nucleotide substitution models are widely used in molecular evolution\cite{Yang2006,Graur1999}
for example to deduce distances between sequences. Some of these models
have few parameters or have particular symmetries. For these models,
it is worthwhile to express relation (\ref{eq:dispersion}) for large
times into an even more explicit form and compute the dispersion number
as an explicit function of the parameters. We provide below such a
computation for some of the most commonly used models.

For the K80 model proposed by Kimura\cite{Kimura1980} , all diagonal
elements of the substitution matrix are equal; hence, relation (\ref{eq:dispersion})
implies that $R=1$.

\subsection{T92 model.}

Tamura \cite{TAMURA1992} introduced a two parameter model (T92) extending
the K80 model to take into account biases in G+C contents. Solving
relation (\ref{eq:dispersion}) explicitly for this model, we find
for the dispersion index 
\begin{equation}
R=1+\frac{2k^{2}}{k+1}\frac{\theta(1-\theta)(2\theta-1)^{2}}{1+2k\theta(1-\theta)}\label{eq:Rtamura92}
\end{equation}
Here $k=\alpha/\beta$, where $\alpha$ and $\beta$ are the two parameters
of the original T92 model. A similar expression was found by Zheng\cite{Zheng2001}.
For a given $k$, the maximum value of $R$ is 
\begin{eqnarray*}
R^{*} & = & 1+\frac{\left(\sqrt{2+k}-\sqrt{2}\right)^{2}}{k+1}
\end{eqnarray*}
And it is straightforward to show that in this case 
\[
R\in[1,2]
\]
although even reaching a maximum value for $R=1.5$ will necessitate
strong asymmetries in the substitution rates (such as $k=18.8$ and
$\theta=0.063$).

\subsection{TN93 model.\label{sub:TN93}}

Tamura and Nei\cite{Tamura1993} proposed a generalization of the
F81\cite{Felsenstein1981} and HKY85 \cite{Hasegawa1985} models which
allows for biases in the equilibrium probabilities, different rates
of transition vs transversion and for different rates of transitions.
The corresponding substitution matrix is 
\[
\mathbf{Q}_{\mbox{TN93}}=\mbox{\ensuremath{\mu}}\left(\begin{array}{cccc}
* & k_{1}\pi_{1} & \pi_{1} & \pi_{1}\\
k_{1}\pi_{2} & * & \pi_{2} & \pi_{2}\\
\pi_{3} & \pi_{3} & * & k_{2}\pi_{3}\\
\pi_{4} & \pi_{4} & k_{2}\pi_{4} & *
\end{array}\right)
\]
a specific case of this model where $k_{1}=k_{2}$ corresponds to
the HKY85 model, while $k_{1}=k_{2}=1$ corresponds to that of F81
(also called ``equal input'' ). Solving equation (\ref{eq:dispersion})
leads to 
\begin{equation}
R=1+\frac{2}{\bar{m}}\sum_{i<j}C_{ij}(m^{i}-m^{j})^{2}\label{eq:RTN93}
\end{equation}
where $m^{i}$ are the (negative of) diagonal elements of $\mathbf{Q}$,
$\bar{m}=\sum_{i}m^{i}\pi_{i}$; $C_{ij}$ are defined as 
\begin{eqnarray*}
C_{12} & = & \pi_{1}\pi_{2}\frac{1-(k_{1}-1)(\pi_{3}+\pi_{4})}{1+(k_{1}-1)(\pi_{1}+\pi_{2})}\\
C_{34} & = & \pi_{3}\pi_{4}\frac{1-(k_{2}-1)(\pi_{1}+\pi_{2})}{1+(k_{2}-1)(\pi_{3}+\pi_{4})}\\
C_{ij} & = & \pi_{i}\pi_{j}\,\,\,\,\mbox{for other}\,i,j
\end{eqnarray*}
For the specific case $k_{1}=k_{2}=1$ (equal input or F81 model),
expression (\ref{eq:RTN93}) takes a particularly simple form
\begin{eqnarray}
R & = & 1+\left(\sum_{i<j}\pi_{i}\pi_{j}(\pi_{i}-\pi_{j})^{2}\right)/\left(\sum_{i<j}\pi_{i}\pi_{j}\right)\label{eq:REI1}\\
 & = & 1+2\frac{\sum_{i}\pi_{i}^{3}-\left(\sum_{i}\pi_{i}^{2}\right)^{2}}{1-\sum_{i}\pi_{i}^{2}}\label{eq:REI2}
\end{eqnarray}
One can deduce relation (\ref{eq:REI1}) from (\ref{eq:REI2}) by
noting that $\sum_{i}\pi_{i}=1$. As every term of the first sum in
relation (\ref{eq:REI1}) is smaller than the corresponding term in
the second sum: 
\[
R_{\mbox{F81}}\in[1,2]
\]
The lower bound is reached for $\left|\pi\right\rangle =(1/4)(1,1,1,1)^{T}$,
while the upper bound is reached when one of the $\pi_{i}$ approaches
1. Zheng \cite{Zheng2001} has also computed an expression for the
dispersion index for the F81 model; his solution however is rather
complicated.

For the general TN93, relation $R\le2$ no longer holds. For example,
for $\left|\pi\right\rangle =(0.6-\epsilon,\epsilon,0.2,0.2$), the
dispersion index is 
\[
R_{\mbox{TN}}=0.04+0.24k_{2}+6/(6+k_{2})+O(\epsilon)
\]
and $R$ can become arbitrarily large with appropriate values of $k_{2}$. 

\begin{figure}
\begin{centering}
\includegraphics[width=0.8\columnwidth]{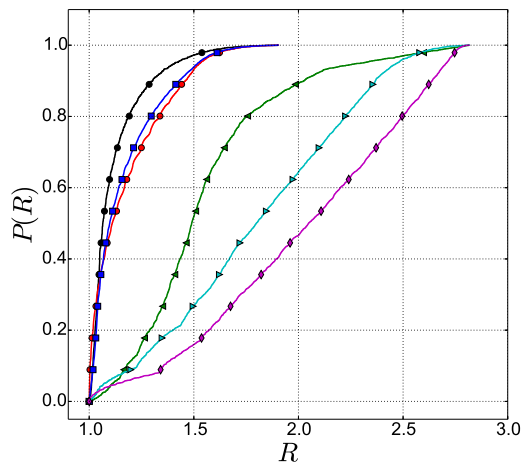}
\par\end{centering}

\caption{Cumulative histogram of the dispersion index $R$ of the TN93 model
and its specific cases. The three dimensional space $\left|\pi\right\rangle =(\pi_{1},\pi_{2},\pi_{3},\pi_{4})$,
$\sum_{i}\pi_{i}=1$ is scanned by steps of $d\pi=0.025$ ($\approx11200$
points). For each value of $\left|\pi\right\rangle $, the dispersion
index of the corresponding substitution matrix $\mathbf{Q}_{\mbox{TN93}}$
is computed from relation (21). Black circles : the F81 model ($k_{1}=k_{2}=1)$;
red diamonds and green left triangles correspond the HKY85 with respectively
$k_{1}=k_{2}=0.1$ and $10$; blue squares, cyan right triangles and
magenta up triangles correspond to TN93 model with $\{k_{1},k_{2}\}=\{0.1,1\},\{1,10\},\{0.1,10\}$
respectively. Permutations of $\{k_{1},k_{2}\}$ lead to the same
results and are not displayed. For each substitution matrix, it has
been checked that solution (\ref{eq:RTN93}) and the general solution
(\ref{eq:dispersion}) are identical. \label{fig:TN93}}

\end{figure}

The simplicity of relation (\ref{eq:RTN93}) allows for the comprehensive
exploration of the hyperplane $\sum_{i}\pi_{i}=1$ , $\pi_{i}>0$.
The results are displayed in figure \ref{fig:TN93} ; to obtain large
values for $R$ such as $R>1.5$ necessitates high asymmetries in
the transition rates and/or strong biases in equilibrium probabilities
of states . 

\section{Statistical investigation of the dispersion index and the influence
of sparseness.\label{sec:generalmodel}}

The relation (\ref{eq:dispersion}) can be solved explicitly for general
substitution matrices. However, a general substitution matrix of dimension
4 has 11 free parameters (substitution matrices are defined up to
a scaling parameter); explicit solution of (\ref{eq:dispersion})
as a function of substitution matrix parameters is rather cumbersome
and does not provide insightful information. 

An exchangeable (time reversible, GTR) substitution matrix has the
additional constraint\cite{Tavare1986} $m_{j}^{i}\pi_{i}=m_{i}^{j}\pi_{j}$.
Considering only exchangeable matrices reduces the number of free
parameters to 9, but the parameter space is still too large to be
explored systematically. 

We can however sample the parameter space by generating a statistically
significant ensemble of substitution matrices $\mathbf{Q}$ and get
an estimate of the probability distribution of the dispersion index
$R$. The simplicity of relation (\ref{eq:dispersion}) allows us
to generate $10^{7}$ random matrices for each class (see section
\ref{sec:Methods}) and compute their associated $R$ in a few minutes
with a usual normal computer : depending on the dimension of $\mathbf{Q}$
(from 4 to 20) this computation takes between 2 and 10 minutes. 

\begin{figure}
\begin{centering}
\includegraphics[width=0.8\columnwidth]{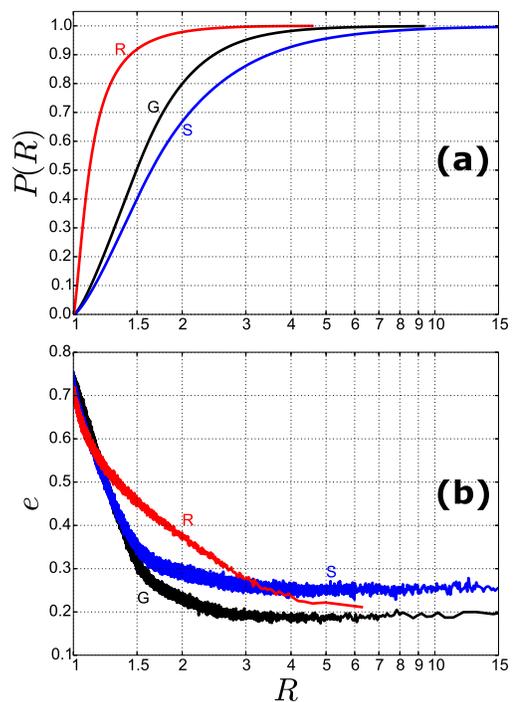}
\par\end{centering}

\caption{Statistical study of $4\times4$ substitution matrices for (i) GTR
matrices (``G'',black curves) ; (ii) arbitrary matrices (``R'',
red curves) and (iii) sparse GTR matrices (``S'', blue curves).
In each case, $10^{7}$ matrices are generated and for each matrix,
its dispersion index $R$ and its eccentricity $e=\mbox{min}(\pi_{i})$
are computed, where $\pi_{i}$ is the equilibrium probability of state
$i$. In each case, the data are sorted by $R$ value to compute the
cumulative histogram (a). figure (b) shows the relation between $e$
and $R$. To make visible the statistical relation between $e$ and
$R$, a moving average of size 1000 data points is applied to the
$10^{7}$sorted $(R,e)$ data in each data set and the result $(R_{m},e_{m})$
is displayed in the lower plot (b). \label{fig:QRstats}}

\end{figure}

Figure \ref{fig:QRstats}.a shows the cumulative probability $P(R)$
for both arbitrary (R) and GTR (G) matrices, computed from $10^{7}$
matrices in each case. We observe that arbitrary matrices produce
statistically low dispersal indices : $P(R>1.5)=0.08$ and $P(R>3)=2.7\times10^{-3}$.
The GTR matrices have statistically higher dispersion indices: $P(R>1.5)=0.495$
and $P(R>3)=0.048$. Still, values larger than $R=5$, as has been
reported in the literature\cite{Cutler2000a}, have a very low probability
($1.4\times10^{-4}$ for random matrices and $7\times10^{-3}$ for
GTR matrices). 

We observed in the preceding section that for each class of matrices,
high values of $R$ are generally associated with large biases in
the equilibrium probabilities, \emph{i.e.} a given state would have
a very low equilibrium probability in order to allow for large $R$.
We can investigate how this observation holds for general and GTR
matrices. For each $K\times K$ matrix $\mathbf{Q}$ that is generated
we quantify its relative eccentricity by 
\[
e=K\,\mbox{min}_{i}(\pi_{i})
\]
The relation between $R$ and $e$ is statistical: matrices with dispersion
index in $[R,R+dR]$ will have a range of $e$ and we display the
average\emph{ $e$} for each small interval (Figure \ref{fig:QRstats}.b).
We observe again that high values of the dispersion index in each
class of matrices requires high bias in equilibrium probabilities
of states.

Another effect that can increase the dispersion index of a matrix
is its sparseness. This effect was investigated by Raval \cite{Raval2007}
for a random walk on neutral networks, which we generalize here. Until
now, we have examined fully connected graphs, \emph{i.e. }substitution
processes where the random variable $X$ can jump from any state $i$
to any other state $j$. For a 4 states random variable, each node
of the connectivity graph\emph{ }is of degree 3 ($d_{G}=3$). This
statement may however be too restrictive. Consider for example a $4\times4$
nucleotide substitution matrix for synonymous substitutions. Depending
on the identity of the codon to which it belongs, a nucleotide can
only mutate to a subset of other nucleotides. For example, for the
third codon of Tyrosine, only $\mbox{T}\leftrightarrow\mbox{C}$ transitions
are allowed, while for the third codon of Alanine, all substitutions
are synonymous. For a given protein sequence, the mean nucleotide
synonymous substitution graph is therefore of degree smaller than
3. In general, the degree of each state (node) $i$ is given by the
number (minus one) of non-zero elements of the $i-$th column in the
associated substitution matrix.

We can investigate the effect of sparseness of substitution matrices
on the dispersion index with the formalism developed above. Figure
\ref{fig:QRstats} shows the probability distribution of $R$ for
GTR matrices with $d_{G}=2$. As it can be observed, the dispersion
index distribution for GTR matrices is shifted to higher values and
$P(R>5)$ increases six fold from 0.007 (for $d_{G}=3)$ to 0.044
(for $d_{G}=2$).

\begin{figure}
\begin{centering}
\includegraphics[width=0.8\columnwidth]{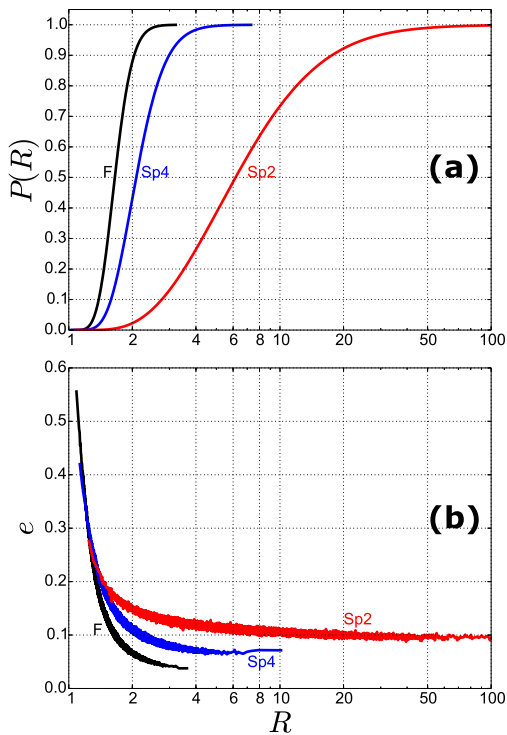}
\par\end{centering}

\caption{Effect of sparseness of $16\times16$ GTR substitution matrices, for
three different connectivities. For each class, $10^{7}$ random matrices
are generated and their statistical properties displayed . Black
lines (marked F) : fully connected graphs $d_{G}=15$ ; blue lines
(marked Sp4) $d_{G}=4$ ; red lines (marked Sp4) $d_{G}=2$. (a) The
cumulative probability $P(R)$ for each class. (b) the mean eccentricity
as a function of dispersion index for each class. \label{fig:sparseK16}}
\end{figure}

The effect of sparseness can be investigated better by considering
higher dimensional substitution matrices. Consider a random variable
$X$ that can take 16 different values. Figure \ref{fig:sparseK16}
shows the effect of sparseness of $\mathbf{Q}$ on the distribution
of the dispersion index. We have considered GTR matrices in three
cases : (i) fully connected transition graphs ($d_{G}=15$) ; (ii)
regular graphs of degree 4 where two different states $i,j=1$...16
are connected if their binary representations are one mutation apart
; and (iii) regular graphs of degree 2. For each class, $10^{7}$
matrices are generated. As can be observed, the sparseness shifts
the dispersion index distribution to the right : the median in the
three cases is respectively 1.65, 2.08 and 6.15. 

\begin{figure}
\begin{centering}
\includegraphics[width=0.8\columnwidth]{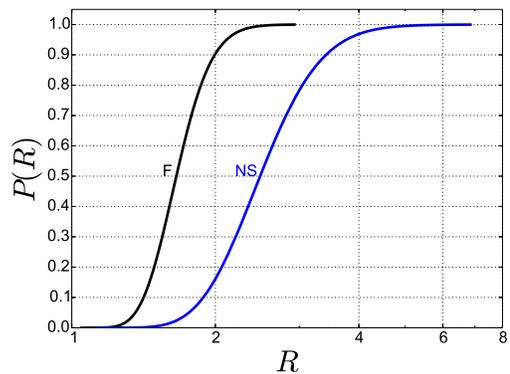}
\par\end{centering}

\caption{Cumulative probability of the dispersion index for Fully connected
(black, F) and non-synonymous (blue, NS) amino acid $20\times20$
substitution matrix. For the fully connected matrix, any amino acid
can be replaced by any other. For the NS matrices, only amino acids
one nucleotide mutation apart can replace each other. For each case,
$10^{7}$ $20\times20$ GTR matrices are generated and their dispersion
index $R$ are computed. For the NS matrices, transitions are weighted
by the number of nucleotide substitutions that can lead from one amino
acid to another: there are for example 6 nucleotide mutations that
transform a Phenylalanine into a Leucine, but only one mutation that
transforms a Lysine into a Isoleucine.\label{fig:AA}}
\end{figure}

A more insightful model would be $20\times20$ GTR matrices for amino
acid substitutions. We compare the case of fully connected graphs
(F) where any amino acid can replace any other one to the case where
only amino acids one nucleotide mutation apart can replace each other
(non-synonymous substitution, NS). The average degree of the graph
in the latter case is $\bar{d}_{G}=7.5$. As before, we generate $10^{7}$
random matrices in each class and compute their statistical properties.
We observe again (Figure \ref{fig:AA}) that the distribution of $R$
is shifted to the right for the NS graphs, where the median is $R_{\mbox{NS}}=2.48$,
compared to $R_{\mbox{F}}=1.66$ for fully connected graphs. 

For specific amino acid substitution matrices used in the literature
such as WAG\cite{Whelan2001}, LG\cite{Le2008} and IDR\cite{Szalkowski2011},
the index of dispersion is 1.253, 1.196 and 1.242 respectively. 

\section{Discussion and conclusion.}

The substitution process (of nucleotides, amino acids, ...) and therefore
the number of substitutions $n$ that take place during time $t$
are stochastic. One of the most fundamental task in molecular evolutionary
investigation is to characterize the random variable $n$ from molecular
data. 

A given model for the substitution process in the form of a substitution
matrix $\mathbf{Q}$ enables us to estimate the \emph{mean} number
of substitution $\left\langle n\right\rangle $ that occur during
a time $t$. The mean depends only on the diagonal elements of $\mathbf{Q}$
and the equilibrium probabilities of states $\pi_{i}$: 
\begin{equation}
\left\langle n(t)\right\rangle =-tQ_{i}^{i}\pi_{i}=-t\mbox{\,tr}(\mathbf{Q}\Pi)\label{eq:disc1}
\end{equation}
where $\mbox{tr() }$designates the trace operator. 

In molecular evolution, the main observable is the probability $p_{d}(t)$
that two different sequences are different at a given site. Denoting
$\mathbf{U}(t)=\exp(t\mathbf{Q})$, and assuming that both sequences
are at equilibrium \cite{Yang2006}, 
\begin{equation}
p_{d}(t)=1-U_{i}^{i}\pi_{i}=1-\mbox{tr}(\mathbf{U}\Pi)\label{eq:disc2}
\end{equation}
One can estimate $p_{d}(t)$ from the fraction of observed differences
between two sequences $\hat{p}$. By eliminating time in relations
(\ref{eq:disc1},\ref{eq:disc2}), it is then possible to relate the
estimators $\hat{d}$ (of $\left\langle n\right\rangle $) and $\hat{p}$
\begin{equation}
\hat{d}=f(\hat{p})\label{eq:disc3}
\end{equation}
For sequences of length $L$, $\hat{p}$ is given by a binomial distribution
$B(L,p)$ and the variance of the distance estimator $\hat{d}$ can
be deduced from relation (\ref{eq:disc3}). This quantity however
is very different from the \emph{intrinsic} variance of the substitution
number.

The mean of substitution number, its estimator $\hat{d}$ and the
variance of the estimator are only the first step in characterizing
a random variable. The next crucial step is to evaluate the variance
$V$ of this number. What we have achieved in this article is to find
a simple expression for $V$. In particular, we have shown that for
both short and long time scales, the variance $V$ can be easily deduced
from $\mathbf{Q}$. For long times, the procedure is similar to deriving
the equilibrium probabilities $\pi_{i}$ from $\mathbf{Q},$\emph{
i.e.} we only need to solve a linear equation associated with $\mathbf{Q}$
(relation \ref{eq:deltaVlong}). For short times, only the diagonal
elements of $\mathbf{Q}$ are required to compute $V$ (relation \ref{eq:shortterm3}). 

A long standing debate in the neutral theory of evolution concerns
the value of dispersion index $R=V/\left\langle n\right\rangle $.
On the one hand, the exact solution of this paper is used to demonstrate
that in general, any substitution process given by a matrix $\mathbf{Q}$
is overdispersed, \emph{i.e.} $R\ge1$, and the equality can be observed
only for trivial models where all diagonal elements of $\mathbf{Q}$
are equal. On the other hand, comprehensive investigation of various
substitution models (section \ref{sec:specificmodel},\ref{sec:generalmodel})
shows that models that produce $R$ much larger than $\approx2$ generally
require strong biases in the equilibrium probabilities of states.
One possibility to produce a higher dispersion index is sparse matrices,
where the ensemble of possible transitions has been reduced. 

The substitution models we have considered here can be applied to
sequences where the $\mathbf{Q}$ matrix is the same for all sites.
For nucleotide sequences, this can describe the evolution of non-coding
sequences or synonymous substitutions (by taking into account the
sparseness of the matrix). On the other hand, for amino acid substitution,
it is well known that some sites are nearly constant or evolve at
a slower pace than other sites. A natural extension of the present
work would be to take into account substitution matrices that vary
among sites, drawing for example their scaling factor from a Gamma
distribution \cite{Yang2006}. The formalism we have developed in
this article can be readily adapted to such an extension and the variance
of the substitution number can be computed for variable substitution
matrices.

\section{Methods.\label{sec:Methods}}

Numerical simulation of stochastic equations use the Gillespie algorithm
\cite{Gillespie1977} and are written in C++ language. To compute
the dispersion index of a given matrix $\mathbf{Q}$, we generate
$10^{6}$ random path over a time period of 1000. To compare the analytical
solutions given in this article (figure \ref{fig:ThSim}) to stochastic
simulations (figure \ref{fig:ThSim}), we generated $3.6\times10^{5}$
random matrices and numerically computed their dispersion index by
the above method. This numerical simulation took approximately 10
days on a 60 core cluster. 

All linear algebra numerical computations and all data processing
were performed with the high-level Julia language \cite{Bezanson2014}.
Computing the analytical dispersion index for the above $3.6\times10^{5}$
random matrices took about 5 seconds on a normal desktop computer,
using only one core. 

To generate random GTR matrices, we use the factorization $\mathbf{Q}=\mathbf{S}\Pi^{-1}$
(see appendix \ref{sec:RL1}) which allows for independent generation
of the $K(K-1)/2$ elements of the symmetric matrix $\mathbf{S}$
and $K-1$ elements of $\Pi$. For arbitrary matrices, we draw the
$K(K-1)$ elements of the matrix. All random generator used in this
work are uniform(0,1).

\paragraph{Acknowledgments.}

We are grateful to Erik Geissler, Olivier Rivoire and Ivan Junier
for fruitful discussions and critical reading of the manuscript.

\appendix

\section{Mean and variance equation.\label{sec:Mean-and-variance}}

Obtaining equations of the moments such as (\ref{eq:nmoment1},\ref{eq:nmoment2})
from the Master equation is a standard procedure of stochastic processes\cite{Gardiner2004,Houchmandzadeh2009a}.
We give here the outline of the derivation.

Consider the master equation (\ref{eq:MasterN}) 
\begin{equation}
\frac{d}{dt}\left|p^{n}\right>=-\mathbf{D}\left|p^{n}\right>+\mathbf{M}\left|p^{n-1}\right>\label{eq:masterNprime}
\end{equation}
which is a system of $K$ equations for the $p_{i}^{n}(t)$, written
in vectorial form. Multiplying each row by $n$ and summing over all
$n$ leads, in vectorial form, to 
\[
\frac{d}{dt}\sum_{n}\left|np^{n}\right>=-\mathbf{D}\sum_{n}\left|np^{n}\right>+\mathbf{M}\sum_{n}\left|np^{n-1}\right>
\]
The term $\sum_{n}\left|np^{n}\right>$ was defined as the vector
of the partial means $\left|n\right>$. For the second term, we have
\begin{equation}
\sum_{n}\left|np^{n-1}\right>=\sum_{n}\left|(n+1)p^{n}\right>=\left|n\right>+\left|p\right>\label{eq:chgvar}
\end{equation}
where $\left|p\right>=\sum_{n}\left|p^{n}\right>$ is the probability
density of the random variable $X$ being in state $i$, whose dynamics
is given by relation (\ref{eq:mainmasterP}). For the initial condition
$\left|p(0)\right>=\left|\pi\right>$, we have at all times 
\[
\left|p(t)\right>=\left|\pi\right>
\]
so the moment equation is 
\[
\frac{d}{dt}\left|n\right>=(-\mathbf{D}+\mathbf{M})\left|n\right>+\mathbf{M}\left|\pi\right>
\]
 which is relation (\ref{eq:moment1}). Note that by definition, $\mathbf{M}\left|\pi\right>=\mathbf{D}\left|\pi\right>$. 

The equation for the second moment (\ref{eq:nmoment2}) is obtained
by the same procedure where each row of equation (\ref{eq:masterNprime})
is multiplied by $n^{2}$. Higher moments and the probability generating
function equations can be obtained by similar computations.

\section{Proof of over dispersion for GTR substitution matrices.\label{sec:RL1}}

As we have seen in relation (\ref{eq:dispersion}), the dispersion
index for long times is 
\[
R=1+2\frac{\left\langle m|r\right\rangle }{\bar{m}}
\]
We must demonstrate that $\left\langle m|r\right\rangle \ge0$ to
prove that $R\ge1$. We give here the proof for GTR matrices. These
matrices can be factorized into 
\begin{equation}
\mathbf{Q}=\mathbf{S}.\mathbf{\Pi^{-1}}\label{eq:GTR}
\end{equation}
where $\Pi=\mbox{diag}(\pi_{1},...,\pi_{k})$ and $\mathbf{S}$ is
a symmetric matrix of positive non-diagonal elements whose columns
(and rows) sum to zero. Note that in the literature\cite{Yang2006},
a slightly different factorization is used in the form of $\mathbf{Q}=\Pi.\mathbf{F}$,
where $\mathbf{F}$ is a symmetric matrix (we stress again that in
our notation, the substitution matrix is the transpose of that used
in most of the literature). The advantage of the factorization (\ref{eq:GTR})
is that except for one zero eigenvalue, all other eigenvalues of $\mathbf{S}$
are negative. $\mathbf{S}$ can be therefore be written as \textbf{
\begin{equation}
\mathbf{S}=\sum_{i=2}^{K}\lambda_{i}\left|v_{i}\right\rangle \left\langle v_{i}\right|\label{eq:Spropre}
\end{equation}
}where $\left|v_{i}\right\rangle $ and $\left\langle v_{i}\right|$
are the right and left orthonormal eigenvectors of $\mathbf{S}$ associated
with the eigenvalue $\lambda_{i}$. The pseudo-inverse of $\mathbf{S}$
is defined as 
\begin{equation}
\tilde{\mathbf{S}}^{-1}=\sum_{i=2}^{K}\lambda_{i}^{-1}\left|v_{i}\right\rangle \left\langle v_{i}\right|\label{eq:Smoins1}
\end{equation}
and it is strictly negative definite . 

The vector $\left|r\right\rangle $ is the solution of the linear
equation 
\begin{eqnarray}
\mathbf{Q}\left|r\right\rangle  & = & \left|h\right\rangle \label{eq:qrh1}\\
\left\langle 1|r\right\rangle  & = & 0\label{eq:qrh2}
\end{eqnarray}
where $h_{i}=(\bar{m}-m^{i})\pi_{i}$. The general solution of the
undetermined equation (\ref{eq:qrh1}) is therefore 
\[
\left|r\right\rangle =C\left|\pi\right\rangle +\Pi\tilde{\mathbf{S}}^{-1}\left|h\right\rangle 
\]
where the constant $C$ is determined from the condition (\ref{eq:qrh2}).
On the other hand 
\begin{eqnarray*}
\left\langle m\right| & = & \left\langle m\right|-\bar{m}\left\langle 1\right|+\bar{m}\left\langle 1\right|\\
 & = & -\left\langle h\right|\Pi^{-1}+\bar{m}\left\langle 1\right|
\end{eqnarray*}
And thus 
\begin{eqnarray}
\left\langle m|r\right\rangle  & = & -\left\langle h|\Pi^{-1}|r\right\rangle +\bar{m}\left\langle 1|r\right\rangle \nonumber \\
 & = & -\left\langle h|\tilde{\mathbf{S}}^{-1}|h\right\rangle -C\left\langle h|1\right\rangle \nonumber \\
 & = & -\left\langle h|\tilde{\mathbf{S}}^{-1}|h\right\rangle \label{eq:quadraticform}
\end{eqnarray}
where we have used the fact that $\left\langle 1|r\right\rangle =\left\langle 1|h\right\rangle =0$.
As $\tilde{\mathbf{S}}^{-1}$ is negative definite , 
\[
\left\langle m|r\right\rangle \ge0
\]
Moreover, the equality is reached only when $\left|h\right\rangle =\left|0\right\rangle $,\emph{
i.e.} for $\pi_{i}\ne0$, only when all diagonal elements of $\mathbf{Q}$
are equal. To see this, we can expand relation (\ref{eq:quadraticform})
\[
\left\langle m|r\right\rangle =-\sum_{i=2}^{K}\lambda_{i}^{-1}\left\langle v_{i}|h\right\rangle ^{2}
\]
The only way to obtain $\left\langle m|r\right\rangle =0$ is to have
$\left\langle v_{i}|h\right\rangle =0$ for $i=2..K$. As on the other
hand, $\left\langle v_{1}|h\right\rangle =\left\langle 1|h\right\rangle =0$
we must have $\left|h\right\rangle =\left|0\right\rangle $.

\section{Dispersion index for all times.\label{sec:alltimes}}

In subsection \ref{sub:Solution} we gave the long (eq. \ref{eq:dispersion})
and short (eq.\ref{eq:shorttimevar}) time solution of the variance.
For all the specific models used in the literature (section \ref{sec:specificmodel}),
the variance at all times can also be determined explicitly through
relation (\ref{eq:variance})
\begin{equation}
\delta V=2\left\langle \tilde{m}|\left(It+\tilde{\mathbf{Q}}^{-1}(\mathbf{I}-e^{\tilde{\mathbf{Q}}t})\right)\tilde{\mathbf{Q}}^{-1}|\left|\tilde{h}\right\rangle \right\rangle \label{eq:deltaV2}
\end{equation}
The procedure requires the computation of $\exp(\mathbf{Q}t)$ and
is analogous to the determination of $\left\langle n\right\rangle $
from sequence dissimilarities\cite{Yang2006,Zheng2001}. 

As an example, consider the equal input model (F81) which we studied
in subsection \ref{sub:TN93}. For this model, the reduced matrix
is simply 
\[
\tilde{\mathbf{Q}}=-\mu\mathbf{I}_{3}
\]
where $\mbox{\textbf{I}}_{3}$ is the $3\times3$ identity matrix
and therefore $\exp(\tilde{\mathbf{Q}}t)=\exp(-\mu t)\mathbf{I}_{3}$.
Relation (\ref{eq:deltaV2}) then becomes 
\begin{eqnarray*}
\delta V & =- & \frac{2}{\mu^{2}}(-1+\mu t+e^{-\mu t})\left\langle \tilde{m}|\tilde{h}\right\rangle 
\end{eqnarray*}
We have previously shown (eq. \ref{eq:shortterm3}) that generally
\[
-\left\langle \tilde{m}|\tilde{h}\right\rangle =\sum_{i=1}^{K}(\bar{m}-m^{i})^{2}\pi_{i}=v_{m}
\]
and for the F81 model, 
\[
\mu^{-2}v_{m}=\sum_{i=1}^{K}\pi_{i}^{3}-\left(\sum_{i=1}^{K}\pi_{i}^{2}\right)^{2}
\]
However, the time can be expressed as a function of mean the substitution
number. Finally, for the F80 model, and setting $\mu=1$ without loss
of generality, the dispersion index for all times is 
\[
R(\left\langle n\right\rangle )=1+2\left(\frac{\left\langle n\right\rangle }{\bar{m}}+e^{-\left\langle n\right\rangle /\bar{m}}-1\right)\frac{v_{m}}{\left\langle n\right\rangle }
\]

\bibliographystyle{unsrt}
\bibliography{Dispersion_Index}

\begin{thebibliography}{10}

\bibitem{Graur1999}
Dan Graur and Wen-Hsiung Li.
\newblock {\em {Fundamentals of Molecular Evolution}}.
\newblock Sinauer Associates Inc.,U.S, 1999.

\bibitem{Yang2006}
Ziheng Yang.
\newblock {Computational molecular evolution}.
\newblock {\em Oxford series in ecology and evolution}, 2006.

\bibitem{Kimura1984}
Motoo Kimura.
\newblock {\em {The Neutral Theory of Molecular Evolution}}.
\newblock Cambridge University Press, 1984.

\bibitem{Bromham2003}
Lindell Bromham and David Penny.
\newblock {The modern molecular clock.}
\newblock {\em Nature reviews. Genetics}, 4(3):216--24, mar 2003.

\bibitem{Ho2014}
Simon Y~W Ho and Sebasti{\'{a}}n Duch{\^{e}}ne.
\newblock {Molecular-clock methods for estimating evolutionary rates and
  timescales.}
\newblock {\em Molecular ecology}, 23(24):5947--65, dec 2014.

\bibitem{Cutler2000a}
D~J Cutler.
\newblock {The index of dispersion of molecular evolution: slow fluctuations.}
\newblock {\em Theoretical population biology}, 57(2):177--86, mar 2000.

\bibitem{Takahata1991a}
Naoyuki Takahata.
\newblock {Statistical models of the overdispersed molecular clock}.
\newblock {\em Theoretical Population Biology}, 39(3):329--344, jun 1991.

\bibitem{Zheng2001}
Qi~Zheng.
\newblock {On the dispersion index of a Markovian molecular clock}.
\newblock {\em Mathematical Biosciences}, 172(2):115--128, aug 2001.

\bibitem{Bastolla2002}
Ugo Bastolla, Markus Porto, H~Eduardo Roman, and Michele Vendruscolo.
\newblock {Lack of self-averaging in neutral evolution of proteins.}
\newblock {\em Physical review letters}, 89(20):208101, nov 2002.

\bibitem{Wilke2004}
Claus~O Wilke.
\newblock {Molecular clock in neutral protein evolution.}
\newblock {\em BMC genetics}, 5(1):25, aug 2004.

\bibitem{Bloom2007}
Jesse~D Bloom, Alpan Raval, and Claus~O Wilke.
\newblock {Thermodynamics of neutral protein evolution.}
\newblock {\em Genetics}, 175(1):255--66, jan 2007.

\bibitem{Raval2007}
Alpan Raval.
\newblock {Molecular clock on a neutral network.}
\newblock {\em Physical review letters}, 99(13):138104, sep 2007.

\bibitem{Huynen1996}
M.~A. Huynen, P.~F. Stadler, and W.~Fontana.
\newblock {Smoothness within ruggedness: the role of neutrality in adaptation.}
\newblock {\em Proceedings of the National Academy of Sciences},
  93(1):397--401, jan 1996.

\bibitem{Bornberg-Bauer1999}
E.~Bornberg-Bauer and H.~S. Chan.
\newblock {Modeling evolutionary landscapes: Mutational stability, topology,
  and superfunnels in sequence space}.
\newblock {\em Proceedings of the National Academy of Sciences},
  96(19):10689--10694, sep 1999.

\bibitem{VanNimwegen1999}
E.~van Nimwegen, J.~P. Crutchfield, and M.~Huynen.
\newblock {Neutral evolution of mutational robustness}.
\newblock {\em Proceedings of the National Academy of Sciences},
  96(17):9716--9720, aug 1999.

\bibitem{Minin2008}
Vladimir~N Minin and Marc~A Suchard.
\newblock {Counting labeled transitions in continuous-time Markov models of
  evolution.}
\newblock {\em Journal of mathematical biology}, 56(3):391--412, mar 2008.

\bibitem{Kimura1980}
Motoo Kimura.
\newblock {A simple method for estimating evolutionary rates of base
  substitutions through comparative studies of nucleotide sequences}.
\newblock {\em Journal of Molecular Evolution}, 16(2):111--120, jun 1980.

\bibitem{TAMURA1992}
K~Tamura.
\newblock {Estimation of the number of nucleotide substitutions when there are
  strong transition-transversion and G+C-content biases}.
\newblock {\em Molecular biology and evolution}, 9(4):678--687, jul 1992.

\bibitem{Tamura1993}
K~Tamura and M~Nei.
\newblock {Estimation of the number of nucleotide substitutions in the control
  region of mitochondrial DNA in humans and chimpanzees}.
\newblock {\em Mol. Biol. Evol.}, 10(3):512--526, may 1993.

\bibitem{Felsenstein1981}
Joseph Felsenstein.
\newblock {Evolutionary trees from DNA sequences: A maximum likelihood
  approach}.
\newblock {\em Journal of Molecular Evolution}, 17(6):368--376, nov 1981.

\bibitem{Hasegawa1985}
M~Hasegawa, H~Kishino, and T~Yano.
\newblock {Dating of the human-ape splitting by a molecular clock of
  mitochondrial DNA.}
\newblock {\em Journal of molecular evolution}, 22(2):160--174, 1985.

\bibitem{Tavare1986}
Simon Tavar{\'{e}}.
\newblock {Some Probabilistic and Statistical Problems in the Analysis of DNA
  Sequences.}
\newblock {\em Lectures in Mathematics in the Life Sciences}, 17:57, 1986.

\bibitem{Whelan2001}
S.~Whelan and N.~Goldman.
\newblock {A General Empirical Model of Protein Evolution Derived from Multiple
  Protein Families Using a Maximum-Likelihood Approach}.
\newblock {\em Molecular Biology and Evolution}, 18(5):691--699, may 2001.

\bibitem{Le2008}
Si~Quang Le and Olivier Gascuel.
\newblock {An improved general amino acid replacement matrix.}
\newblock {\em Molecular biology and evolution}, 25(7):1307--20, jul 2008.

\bibitem{Szalkowski2011}
Adam~M Szalkowski and Maria Anisimova.
\newblock {Markov models of amino acid substitution to study proteins with
  intrinsically disordered regions.}
\newblock {\em PloS one}, 6(5):e20488, jan 2011.

\bibitem{Gillespie1977}
Daniel~T Gillespie.
\newblock {Exact stochastic simulation of coupled chemical reactions}.
\newblock {\em The Journal of Physical Chemistry}, 81(25):2340--2361, 1977.

\bibitem{Bezanson2014}
Jeff Bezanson, Alan Edelman, Stefan Karpinski, and Viral~B. Shah.
\newblock {Julia: A Fresh Approach to Numerical Computing}.
\newblock page~37, nov 2014.

\bibitem{Gardiner2004}
C~Gardiner.
\newblock {\em {Handbook of Stochastic Methods: for Physics, Chemistry and the
  Natural Sciences}}.
\newblock Springer, 2004.

\bibitem{Houchmandzadeh2009a}
Bahram Houchmandzadeh.
\newblock {Theory of neutral clustering for growing populations}.
\newblock {\em Physical Review E}, 80(5):051920, nov 2009.

\end{thebibliography}

\end{document}